\documentclass[twocolumn,amsmath,amssymb,floatfix,prl,superscriptaddress,showpacs]{revtex4}
\usepackage{amsmath,amssymb,natbib,bm,graphicx,url,epsfig,psfrag}
\usepackage[ansinew]{inputenc}
\usepackage{color}

\begin{document}
\title{How to Directly Measure Kondo Cloud's Length}
\author{Jinhong Park}
\affiliation{Department of Physics, Korea Advanced Institute of
Science and Technology, Daejeon 305-701, Korea}
\author{S.-S. B. Lee}
\affiliation{Department of Physics, Korea Advanced Institute of
Science and Technology, Daejeon 305-701, Korea}
\author{Yuval Oreg}
\affiliation{Department of Condensed Matter Physics, Weizmann
Institute of Science, Rehovot 76100, Israel}
\author{H.-S. Sim} \email{hssim@kaist.ac.kr}
\affiliation{Department of Physics, Korea Advanced Institute of
Science and Technology, Daejeon 305-701, Korea}
\date{\today}
\begin{abstract}
We propose a method to directly measure, by electrical means, the Kondo screening cloud formed by an Anderson impurity coupled to semi-infinite quantum wires, on which an electrostatic gate voltage is applied at distance $L$ from the impurity. We show that the Kondo cloud, and hence the Kondo temperature and the electron conductance through the impurity, are affected by the gate voltage, as $L$ decreases below the Kondo cloud length. Based on this behavior, the cloud length can be experimentally identified by changing $L$ with a keyboard type of gate voltages or tuning the coupling strength between the impurity and the wires.
\end{abstract}

\maketitle


\emph{Introduction.}--- The Kondo effect is a central many-body problem of condensed matter physics~\cite{Kondo,Hewson}. It involves a spin singlet, formed by the spin-spin interaction between a magnetic impurity and surrounding conduction electrons. Deeper understanding of the effect has been achieved by using a quantum dot, that hosts a magnetic impurity spin, under systematic control~\cite{Goldhaber-Gordon,Cronenwett,kouwenhoven,Glazman}. 

Although the Kondo effect is well known,
its spatial features still remain to be addressed. The Kondo spin singlet is formed below the energy scale of Kondo temperature $T_K$. This implies that the singlet is spatially formed over a conduction-electron region of length scale $\xi_K = \hbar v_F / (k_B T_K)$; when $T_K ~\sim 1$ K and the Fermi velocity $v_F \sim 10^5$-$10^6 \, \textrm{m}/\textrm{s}$, $\xi_K \sim 1 \, \mu \textrm{m}$.
The region is called the Kondo screening cloud.  
There have been several proposals~\cite{Gubernatis,Barzykin,Borda,Holzner,Simon_persistent,Balseiro,Simon_dot,Hand,Sorensen_knight,Affleck_density,Busser,Mitchell,Yoshii} for ways to detect the cloud.


Despite the proposals, there has been no conclusive measurement supporting the existence of the Kondo cloud~\cite{Boyce,Affleck_review}. The difficulty to detect the cloud arises because it is a spin cloud showing quantum fluctuations with zero averaged spin.
The cloud manifests itself in
the spin-spin correlation~\cite{Gubernatis,Barzykin,Borda,Holzner} between the impurity and the conduction electrons. However it requires measurements of spin dynamics of time scale $\hbar /(k_B T_K)$. STM studies probing local density of states may be useful for detecting the cloud~\cite{Affleck_density,Busser,Mitchell,Bergmann_G}. Recent STM measurements~\cite{Madhavan,Manoharan,Pruser,Fu} show the Kondo effect in the region away from a magnetic impurity, whose spatial extension is however much shorter than $\xi_K$. Another direction is to study a magnetic impurity in a finite-size system~\cite{Simon_persistent,Balseiro,Simon_dot,Hand,Yoshii,Thimm,Bomze}. Because the cloud cannot extend beyond the finite size, the Kondo effect is strongly affected, and suppressed 
when the system is shorter than $\xi_K$. There has been no conclusive experimental detection of $\xi_K$ in this direction~\cite{Bomze}. 

\begin{figure}[b]
\centering\includegraphics[width=0.44\textwidth]{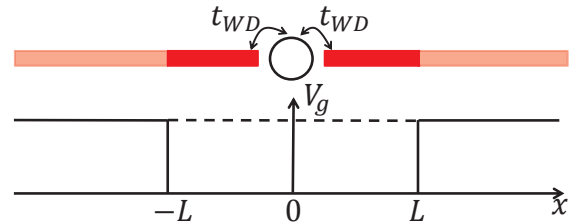}
\caption{(Color Online) Setup for detecting the Kondo cloud. A quantum dot located at $x=0$ hosts a magnetic impurity spin. It couples to quantum wires along $\hat{x}$ axis, with electron tunneling amplitude $t_{WD}$. Gate voltage $V_g$ is applied at distance $L$ from the dot (in $|x| > L$). The Kondo effect becomes sensitive to $V_g$, as $L$ decreases below the cloud length.   
}
\label{Setup1}
\end{figure}

\begin{figure}[tpb]
\includegraphics[width=0.4\textwidth]{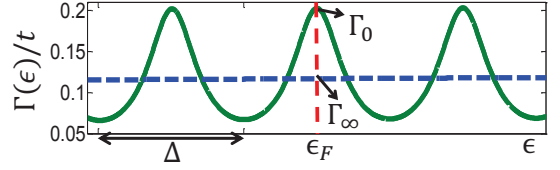}
\caption{(Color Online) Hybridization function $\Gamma(\epsilon)=2\pi t^2_{WD}\rho(\epsilon)$ in unit of $t$ for infinite (blue dashed curve) and finite $L$ (green solid); we choose $\epsilon_0/2t= 0.931$, $t_{WD}/2t = 0.2$, and $eV_g/2t=0.125$. For finite $L$, $\Gamma(\epsilon)$ has resonances with spacing $\Delta \equiv \pi \hbar v_F / L$; $\epsilon_F$ is chosen to be located at a resonance center (red dashed line).
When $L$ is so large that $V_g$ is applied outside the Kondo cloud, the Kondo temperature $T_K = T_{K \infty}$ is determined by $\Gamma_\infty$ and independent of $V_g$. On the other hand, $T_K$ depends on $V_g$ and $L$, when $L$ $\lesssim$ the cloud size $\xi_K$. When $L \ll \xi_K$, $T_K$ is determined by $\Gamma_0 \equiv \Gamma (\epsilon_F)$.
}\label{DensityOfStates2}
\end{figure}

\begin{figure}[tb]
\includegraphics[width=0.49\textwidth]{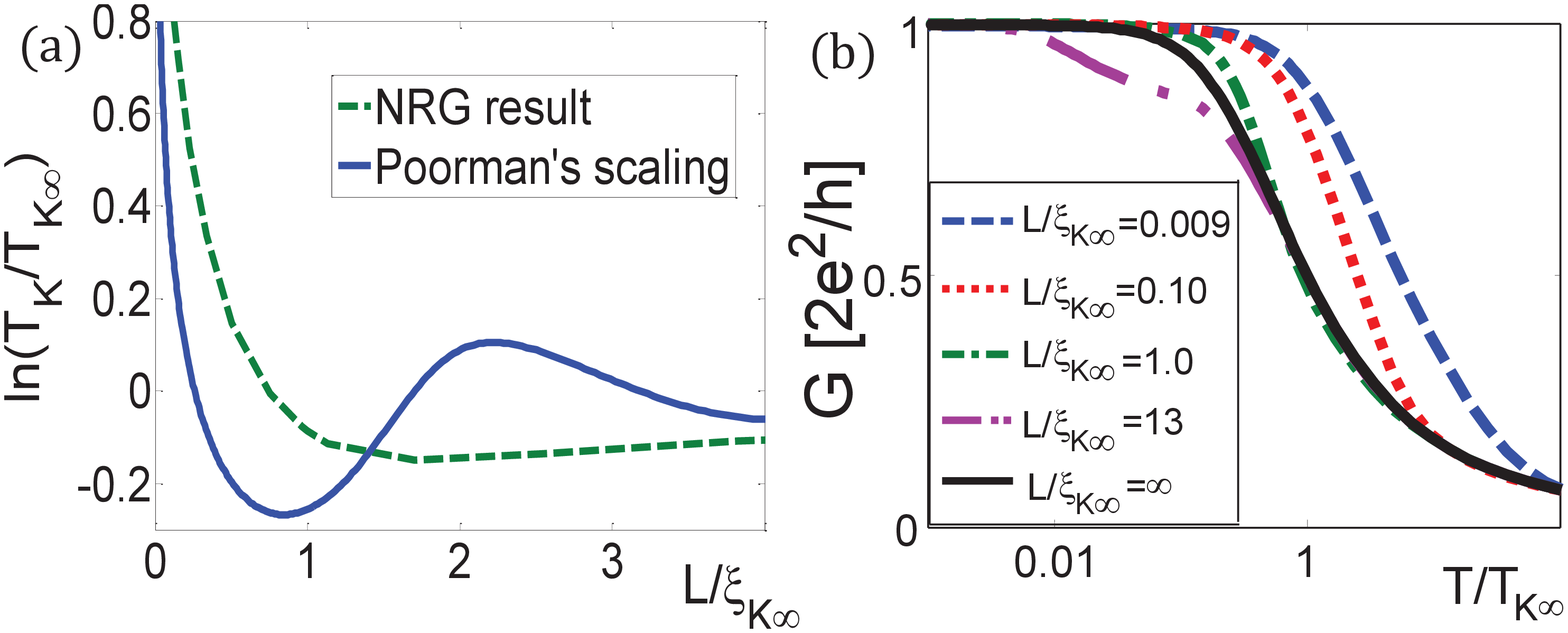}
\caption{ (Color Online) Case A under the resonance condition of $k_F = k_{F,n}$. In this case, one changes $L$ with keeping $t_{WD}$ (hence $T_{K \infty}$ and $\xi_{K \infty}$) constant.
(a) Kondo temperature $T_K$ as a function of $L$, obtained by the poor man scaling (blue solid curve) and NRG (green dashed). The two approaches show qualitatively the same overall behavior that $T_K$ drastically changes for $L \lesssim \xi_{K \infty}$, while $T_K \sim T_{K \infty}$ for $L \gtrsim \xi_{K \infty}$; their discrepancy in $L \gtrsim \xi_{K \infty}$ is discussed in the text.   
(b) NRG result of the temperature $T$ dependence of conductance $G$ for various values of $L/ \xi_{K \infty}$.
We choose $\Gamma_{\infty}/2t=0.28$, $\epsilon_0/2t=0.925$, $eV_g/2t=0.125$, and $U/2t=3.6$.
}\label{CondTkchangingL3}
\end{figure}

\begin{figure}[tb]
\includegraphics[width=0.49\textwidth]{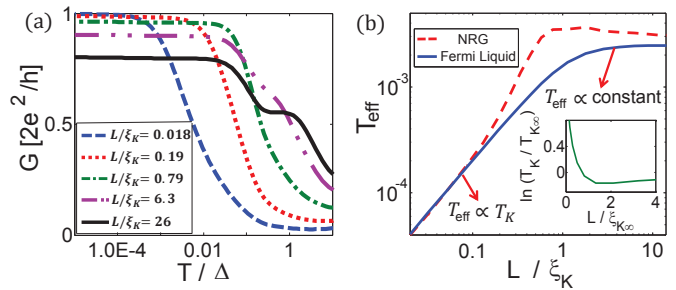}
\caption{ (Color Online) Case B. In this case, one changes $t_{WD}$ (hence $\xi_K$)
with keeping $L$ constant.
(a) The NRG result of $G(T)$ for different values of $t_{WD}$; we show the values of $\xi_K$ instead of $t_{WD}$.   
(b) $T_{\rm eff}(L / \xi_{K})$ [defined in Eq.~\eqref{Teffective}], obtained from the two different approaches of the Fermi liquid theory and the NRG. 
Inset: The NRG result of $T_K (L/ \xi_{K \infty})$ exhibits the same behavior as Fig.~\ref{CondTkchangingL3}(a). We choose $L=100a$, $\epsilon_0/2t=0.931$, $eV_g/2t=0.125$, and $U/2t=3.6$. $\epsilon_F$ is chosen to be located at a center of a resonance of $\Gamma(\epsilon)$, for simplicity.
}\label{CondTkchangingCoup4}
\end{figure}

In this work, we propose a new way of detecting the Kondo cloud, based on the intuition that a change of conduction electrons inside the cloud will affect the Kondo effect. We consider a Kondo impurity formed in a quantum dot coupled to two semi-infinite ballistic quantum wires with electron tunneling amplitude $t_{WD}$ (see Fig.~\ref{Setup1}). Electrostatic gate voltages $V_g$ are applied to the wires (or to only one wire) at distance $L$ from the dot, modifying {\em indirectly} the local density of states $\rho(\epsilon)$ of conduction electrons nearby the dot (Fig.~\ref{DensityOfStates2}). 
We find that $V_g$ does not affect the cloud, when $L \gg \xi_K$. However, when $L \ll \xi_K$, the cloud, hence Kondo temperature $T_K$ and electron conductance $G$ through the dot, are sensitive to $V_g$. The crossover between the two regimes occurs at $L \approx \xi_K$. By measuring $G$ or $T_K$ with varying $L$ or $t_{WD}$ (Figs.~\ref{CondTkchangingL3} and~\ref{CondTkchangingCoup4}), one can detect the crossover and $\xi_K$. We use the poor man scaling~\cite{Anderson_Poorman}, numerical renormalization group study (NRG)~\cite{Wilson,Bulla,Delft}, and Fermi liquid theory~\cite{Nozieres}.

\emph{The setup.}--- 
We describe the wires by the tight-binding Hamiltonian with sites $j$'s in wire $i=l,r$,
\begin{eqnarray} \label{wireHamiltonian}
H_W&=& \sum_{i=l,r}
\sum_{j=1}^{\infty} \sum_{\sigma=\uparrow,\downarrow}
[\epsilon_0 n_{ij\sigma} + (-tc_{ij\sigma}^{\dagger}c_{i(j+1)\sigma}+\text{H.c.})] \nonumber \\
& & - eV_g \sum_{i=l,r} \sum_{j=N+1}^{\infty} \sum_{\sigma=\uparrow,\downarrow}  n_{ij\sigma},
\end{eqnarray}
where $c_{ij\sigma}^{\dagger}$ creates an electron with spin $\sigma$ and energy $\epsilon_0$ at site $j$ in wire $i$, $n_{ij \sigma} \equiv c_{ij\sigma}^{\dagger} c_{ij\sigma}$, 
and $t$ is the hopping energy. 
The last term describes the gate voltage $V_g$ applied in $|x| > L = Na$, where $a$ is the lattice spacing. 
$V_g(x)$ changes at $x=L$ abruptly over the length shorter than the Fermi wave length. 
The dot Hamiltonian $H_D$ is modeled by the Anderson impurity~\cite{Anderson}, $H_D=\sum_{\sigma=\uparrow,\downarrow}\epsilon_{d}d_{\sigma}^{\dagger}d_{\sigma}+Un_{d\uparrow}n_{d\downarrow}$,
where $d_\sigma^\dagger$ creates an electron with energy $\epsilon_d$ and spin $\sigma$ in the dot, $n_{d \sigma} \equiv d^\dagger_\sigma d_\sigma$, and $U$ is the electron repulsive interaction.
$H_T=-t_{WD} \sum_i \sum_{\sigma=\uparrow,\downarrow}(c_{i1\sigma}^{\dagger}d_{\sigma}+\text{H.c.})$ describes electron tunneling between the wire and the dot.

The dot is occupied by a single electron in the Coulomb blockade regime of $\epsilon_d<\epsilon_F$ and $\Gamma (\epsilon_F) \ll-\epsilon_d+\epsilon_F, \,\, U+\epsilon_d-\epsilon_F$, where $\Gamma (\epsilon) = 2 \pi |t_{WD}|^2 \rho(\epsilon)$ is the hybridization function between the dot and the wires, $\rho(\epsilon)$ is the local density of states at energy $\epsilon$ in the neighboring sites $j=1$ of the dot, and $\epsilon_F$ is the Fermi energy.  In this regime, the total Hamiltonian of the setup 
becomes~\cite{Schrieffer,Hewson,Glazman}
\begin{eqnarray}
H = H_D + H_W + H_T \simeq J \vec{s} \cdot \vec{S} + V \sum_{i,\sigma} n_{i1 \sigma} + H_W.
\label{JKHamiltonian}
\end{eqnarray}
Here, the Kondo impurity spin, $\vec{S} = \sum_{\sigma, \sigma'} d_{\sigma}^{\dagger} \vec{\sigma}_{\sigma \sigma'} d_{\sigma'}/2$, couples with the spin of the neighboring conduction electrons, $\vec{s} = \sum_{\sigma, \sigma'} (c_{l1 \sigma}^{\dagger}+c_{r1 \sigma}^\dagger) \vec{\sigma}_{\sigma \sigma'} (c_{l 1 \sigma'} + c_{r 1 \sigma'})/2$, with strength $J = 2 t_{WD}^2 [- 1 /(\epsilon_d - \epsilon_F) + 1 / (U + \epsilon_d-\epsilon_F)]$. The second term describes the potential scattering with strength 
$V = t_{WD}^2 [- 1 / (\epsilon_d-\epsilon_F) - 1 / (U + \epsilon_d-\epsilon_F)]/2$.

\emph{Local density of states.}--- We show how the gate voltage $V_g$ changes the local density of states $\rho (\epsilon)$ at sites $j=1$ of the wires. 
We calculate $\rho(\epsilon)$, by matching the single-particle wavefunctions of $H_W$ between $j=N$ and $N+1$,
\begin{equation} \label{densityofstates}
\rho(\epsilon)=\frac{\sin(qa)\sin^2(ka)}{\pi at [\sin^2(ka)+\frac{eV_g}{t}
\sin[k(N+1)a]\sin(kNa)]}.
\end{equation}
$k$ and $q$ are the wavevectors in $|x| < L$ and $|x| > L$, respectively, satisfying
$\epsilon=\epsilon_0-2t\cos(ka)=\epsilon_0-eV_g-2t\cos(qa)$.
$\Gamma(\epsilon)=2\pi t_{WD}^2 \rho(\epsilon)$ shows resonances with level spacing $\Delta \equiv \pi \hbar v_F / L$; see  Fig.~\ref{DensityOfStates2}. The oscillation amplitude of $\Gamma(\epsilon)$ is proportional to $V_g$. 
The value of $\Gamma(\epsilon_F)$ of the $L \to \infty$ limit is denoted as $\Gamma_\infty$, which equals
the average of $\Gamma(\epsilon)$ around $\epsilon_F$ for finite $L$.


We sketch how the change of $\rho(\epsilon)$ by $V_g$ affects the Kondo effect in different regimes of $L$.
For $L \gg \xi_{K \infty}$ (i.e., $T_{K \infty} \gg \Delta$), the Kondo temperature is determined by $\Gamma_\infty$ as $T_{K \infty} \sim \sqrt{\Gamma_\infty U/2} \exp[ \pi(\epsilon_d-\epsilon_F)(U+\epsilon_d-\epsilon_F) / (2\Gamma_{\infty}U)]$, and the cloud size is $\xi_{K \infty} = \hbar v_F / (k_B T_{K \infty})$. In this regime, the average $\Gamma_{\infty}$ of many resonances of $\Gamma(\epsilon)$ around $\epsilon_F$ determines the Kondo effect, insensitively to $V_g$. 
On the other hand, for $L \ll \xi_{K \infty}$ (i.e., $T_{K \infty} \ll \Delta$), the resonance of $\Gamma(\epsilon)$ located at $\epsilon_F$, namely $\Gamma_0 \equiv \Gamma (\epsilon_F)$,  determines the Kondo effect, resulting in 
$T_{K} = T_{K0} \sim \sqrt{\Gamma_0 U/2} \exp[ \pi(\epsilon_d-\epsilon_F)(U+\epsilon_d-\epsilon_F) / (2\Gamma_0 U)]$ and
$\xi_K = \hbar v_F / (k_B T_{K 0})$.
In this case, $V_g$ affects conduction electrons within the cloud, and modifies $T_K$.

We will discuss how $T_K$ changes between $T_{K0}$ and $T_{K \infty}$ as a function of $L / \xi_K$ in the two possible situations, case A where one changes $L$ with keeping $t_{WD}$ constant, and case B where $t_{WD}$ changes and $L$ remains constant.



\emph{Kondo temperature.}--- We compute $T_K$, using the poor man scaling and the NRG~\cite{Delft}.

In the poor man scaling, the renormalization of 
$J \to J + J^2  (\int_{-D_0}^{-D}+\int_{D}^{D_0})d\epsilon \rho (\epsilon)/|\epsilon|$ is performed with reducing the energy bandwidth of the wire from $D_0$ to $D$, and stopped at the bandwidth where $J^2 \int d\epsilon \rho (\epsilon)/|\epsilon|$ is comparable with $J$. The final bandwidth provides $T_K$,
\begin{eqnarray}
\ln(\frac{T_K}{T_{K\infty}}) & \simeq & -  \frac{eV_g\cos(k_F(2L+a))}{2t\sin^2(k_Fa)}\text{Ci}(\frac{2L}{\xi_{K}}),
\label{scalingbehaviorOfKondotemp} \\
 & \simeq & -  \frac{eV_g}{2t \sin^2 (k_F a)}\text{Ci}(\frac{2L}{\xi_{K}}) \,\,\,\,\,\, \textrm{for} \,\,  k_F = k_{F,n}, \,\,\,\,\,\,
\label{scalingbehaviorOfKondotemp2}
\end{eqnarray}
where $\text{Ci}(y)\equiv \int_{-\infty}^{-y}dy' (\cos y') / y'$, $k_{F,n} = 2 \pi n / (2L +a)$, and $n$ is an integer. 
Equation~\eqref{scalingbehaviorOfKondotemp} is obtained by putting $k \to k_F \equiv k(\epsilon_F)$, $q \to q(\epsilon_F)$, $L \gg a$, $v_F \equiv \frac{1}{\hbar}\frac{\partial\epsilon}{\partial k}\arrowvert_{k_F}= (2at/\hbar)\sin(k_Fa)$, $|eV_g|\ll 2t\sin^2(k_Fa)$, and the linearization of $\epsilon \simeq \epsilon_F + \hbar v_F (k - k_F)$ into Eq.~\eqref{densityofstates}; $k \to k_F$ and $q \to q(\epsilon_F)$ are valid within the small energy scale of $T_K$. We remark that $\xi_{K}$ depends on $L$ and $V_g$ in Eq.~\eqref{scalingbehaviorOfKondotemp}. 

In Eq.~\eqref{scalingbehaviorOfKondotemp}, the term $\text{Ci}(2L/\xi_{K})$ gives the information on the cloud, while another $L$ dependence of the $2 k_F$ oscillation appears because resonance centers in $\Gamma(\epsilon)$ shift across $\epsilon_F$ as $L$ changes. One can focus on the former. 
In case A, where one changes $L$, one can reduce the effect of the $2k_F$ oscillation, by considering the situation that $\epsilon_F$ is located near the bottom of an energy band where the $2k_F$ term slowly oscillates, or by considering the resonance condition of $k_F = k_{F,n}$ where the $2k_F$ term provides the maximum value; see Eq.~\eqref{scalingbehaviorOfKondotemp2}. The resonance condition can be achieved at each value of $L$, by tuning an additional gate voltage applied to the entire region of the wires (not shown in Fig.~\ref{Setup1}) with monitoring the conductance through the dot.
On the other hand, in case B, where one changes $t_{WD}$ with keeping $L$ constant, the term $\cos(k_F(2L+a))$ is constant, hence can be ignored. 

For case A under the resonance situation of $k_F = k_{F,n}$, the poor man scaling in Eq.~\eqref{scalingbehaviorOfKondotemp2} is plotted as a function of $L / \xi_{K \infty}$ (rather than $L / \xi_K$) in Fig.~\ref{CondTkchangingL3}(a). As expected, $T_K$ stays at $T_{K \infty}$ for $\xi_{K \infty} \lesssim L$, drastically changes around $L = \xi_{K \infty}$, and approaches to $T_{K 0 }$ for $L \ll \xi_K$. In addition, the oscillation of $\ln(T_K/T_{K\infty}) \sim \textrm{sinc} (2L /\xi_{K \infty})$ appears for $L \gtrsim \xi_{K \infty}$; we put $\text{Ci}(x) \sim \textrm{sinc} \, x = (\sin x)/x$ and $\xi_K \simeq \xi_{K \infty}$ (valid for $x = 2 L / \xi_{K \infty} \gtrsim 1$) into Eq.~\eqref{scalingbehaviorOfKondotemp}. The oscillation originates from the average effect of $\rho(\epsilon)$ within $T_{K \infty}$, and becomes suppressed for longer $L$ as more ($\sim T_{K \infty} / \Delta$) resonances appear within $T_{K \infty}$.
On the other hand, for $L \ll \xi_{K\infty}$,  we find $\ln (T_K / T_{K \infty}) \propto - \ln (L / \xi_{K})$, using $\text{Ci}(x) \sim \ln(x) + 0.577$ for $x\ll 1$. The above behavior of $T_K (L / \xi_{K \infty})$ reveals the Kondo cloud. 



\emph{Conductance.}--- We compute the temperature $T$ dependence of electron conductance $G$ between the wires through the dot, using the NRG~\cite{Wilson,Bulla,Balseiro,Delft}. 
We will discuss how to extract $\xi_{K \infty}$ from $G(T)$ in cases A and B. 

We continue to discuss case A under the resonance condition of $k_F = k_{F,n}$. In Fig.~\ref{CondTkchangingL3}(b), we plot $G(T)$ for different $L$'s. It is custom~\cite{Goldhaber-Gordon_Kondotemp} to get an estimate for $T_K$ from the temperature at which $G(T)$ equals the half of the zero-temperature conductance $G(T=0, L \to \infty )$ of the $L \to \infty$ case. 
$G(T)$ shows the behavior distinct between $L \gtrsim \xi_{K \infty}$ and $L \lesssim \xi_{K \infty}$. For $L \gtrsim \xi_{K \infty}$, $G(T)$ equals $G(T=0, L \to \infty)/2$ at almost the same temperature, implying that $T_K$ equals $T_{K \infty}$ independent of $L$. On the other hand, for $L \lesssim \xi_{K \infty}$, $G(T)$ shows that $T_K$ changes toward $T_{K0}$ as $L$ decreases. This NRG result agrees with the poorman scaling; see Fig.~\ref{CondTkchangingL3}(a). In this way, one directly measures $\xi_{K \infty}$.

There is a discrepancy between the two curves in Fig.~\ref{CondTkchangingL3}(a). In the NRG case, $\ln(T_K/T_{K\infty})$ decreases only monotonously for $L\gtrsim \xi_K$, without showing the behavior $\textrm{sinc} (2L /\xi_{K \infty})$ of the poor man scaling. The discrepancy may come from the known limitation that the logarithmic discretization scheme of NRG does not perfectly capture the behavior of high-energy states ({higher than} $\Delta$ in our case). 
On the other hand, for $L \ll \xi_{K \infty}$, where low-energy states mainly contribute to the Kondo effect, the NRG shows the same behavior of $\ln (T_K / T_{K \infty}) \sim - \ln (L / \xi_{K})$ as the poorman scaling. 

Next, we discuss case B where one changes $t_{WD}$ with keeping $L$ constant (hence $\Delta = \pi \hbar v_F / L$ is constant). Figure~\ref{CondTkchangingCoup4} shows the NRG result of $G(T)$ for different $\xi_K$'s. We obtain $T_K$ from the high-temperature behavior of $G(T)$ in the same way as above, by choosing the temperature at which $G(T) = G(T = 0, L \to \infty) / 2$. 
The result of $T_K$ agrees with case A; see the inset of Fig.~\ref{CondTkchangingCoup4}(b).
We below suggest another way to see the cloud from the low-temperature behavior of $G(T)$.
Note that as $T$ changes across $\Delta$, $G(T)$ can show a jump due to the resonance structure of $\rho(\epsilon)$, as shown for $L/\xi_K = 26$ in Fig.~\ref{CondTkchangingCoup4}(a).

We describe the regime of $T \ll T_K, \Delta$, using the fixed-point Hamiltonian 
of the Fermi liquid theory~\cite{Nozieres,Glazman,Affleck2},
\begin{eqnarray}
H_{\text{low}} &\simeq& \sum_{k \sigma} \epsilon_{k} c_{k \sigma}^{\dagger}
c_{k \sigma}- \frac{1}{\pi \rho(\epsilon_F)} \sum_{kk'\sigma} (\frac{\epsilon_{k}+\epsilon_{k'}}{2 T_K }+\delta_{p})
 c_{k \sigma}^{\dagger} c_{k' \sigma} \nonumber \\
 && +\frac{1}{\pi T_K \rho^2(\epsilon_F) }\sum_{k_1 k_2 k_3 k_4} c_{k_1 \uparrow}^{\dagger}c_{k_2 \uparrow}c_{k_3 \downarrow}^{\dagger}c_{k_4 \downarrow},
  \label{fixedpointHamiltonian}
\end{eqnarray}
where $c_{k \sigma}^\dagger$ creates an electron with momentum $k$, spin $\sigma$, and energy $\epsilon_k$. $\delta_p$ is the phase shift by the potential scattering, which occurs as the particle-hole symmetry is broken. 
Although $\rho$ depends on $\epsilon$, we take for simplicity $\rho (\epsilon_F)$ in Eq.~\eqref{fixedpointHamiltonian} as a crude approximation.  
The second term of Eq.~\eqref{fixedpointHamiltonian} describes elastic scattering of electrons by the Kondo singlet, with scattering phase shift $\delta(\epsilon)=\pi/2+ (\epsilon-\epsilon_F)/T_K + \delta_p$.
The third term shows repulsive interactions that break the Kondo singlet, 
and contributes to inelastic T-matrix $t^{\text{in}}$ as $-\pi \rho(\epsilon_F) \text{Im} t^{\text{in}}(\epsilon)=[(\epsilon-\epsilon_F)^2+\pi^2 T^2]/(2T_K^2)$. By combining $\delta(\epsilon)$ and $\text{Im} t^{\text{in}}(\epsilon)$, we obtain~\cite{Glazman}
\begin{eqnarray} 
G(T)&=&\frac{2e^2}{h}\frac{1+\cos(2\delta_p)}{2}(1-\frac{\pi^2 T^2}{T_\textrm{eff}^2}),
\label{ConductanceLowTemp} \\
\frac{1}{T_\textrm{eff}^2}&\equiv& \frac{2 \beta}
{3(\alpha + \beta)\Delta^2}+\frac{2\cos(2\delta_p)}{(1+\cos(2\delta_p)){T_K^2}},
\label{Teffective}
\end{eqnarray} 
where $\alpha$ and $\beta$ are constants depending on $V_g$ and $k_F$ but independent of $L$.
$\delta_p$ is obtained by comparing $G(T=0)$ with Eq.~\eqref{ConductanceLowTemp}.
For nonzero $\delta_p$, $G(T=0)$ deviates from the unitary-limit value of $2e^2/h$.
Note that when $\rho$ is independent of $\epsilon$ and the particle-hole symmetry is preserved, $\delta_p =0$ and $\beta = 0$, hence, $T_\textrm{eff} \to T_K$. 

$T_\textrm{eff}$ is obtained by comparing $G(T)$ with Eq.~\eqref{ConductanceLowTemp}
in experiments or in the NRG, while computed from Eq.~\eqref{Teffective} in the Fermi liquid theory. We plot $T_\textrm{eff}(L / \xi_K)$ in Fig.~\ref{CondTkchangingCoup4}(b), showing good agreement between the NRG and the Fermi liquid theory; their quantitative discrepancy may come from our approximation in the Fermi liquid theory.  

The dependence of $T_{\rm eff}$ on $L / \xi_{K}$ or $L / \xi_{K \infty}$ is useful for identifying $\xi_{K}$ in case B, since $\Delta$ is constant so that $T_{\rm eff}(L/\xi_K)$ directly provides the information of $T_K (L / \xi_K)$; see Eq.~\eqref{Teffective}. For $L \gtrsim \xi_K$, $T_\textrm{eff}$ is almost constant, implying that $T_K$ and $\xi_K$ are independent of $L$. 
For $L \lesssim \xi_K$, $T_{\rm eff}$ (hence $T_K$) depends on $L/ \xi_K$. The crossover occurs around $\xi_K \simeq L$. 

 
  

\emph{Discussion.}--- Our proposal may be within experimental reach. Case A, where $L$ varies, may be achieved with keyboard-type gate voltages, while one tunes $t_{WD}$ by a gate in case B. A good candidate for our proposal may be a carbon nanotube, where $T_K \sim 1 \,\, \textrm{K}$ and $\xi_K \sim 1 \mu \textrm{m}$~\cite{Nygard}.

For both the cases, a (single-mode or multi-mode) wire whose Fermi level $\epsilon_F$ lies near the bottom $E_b$ [van Hove singularity (VHS)] of one of the energy bands is useful to achieve a conclusive evidence of the cloud; our results of Eq.~\eqref{scalingbehaviorOfKondotemp} and NRG are applicable to this regime, since they are obtained, taking into account of the energy dependence of $\rho$. In this regime, $\rho(\epsilon)$ is sensitive to $V_g$, hence, it may not be difficult to obtain sizable difference between $T_{K \infty}$ and $T_{K 0}$ 
by $V_g (< \epsilon_F - E_b)$. Our analysis for a single-mode wire is applicable, without modification, to a multi-mode wire, since the band near the VHS governs the Kondo effect dominantly over the other modes.  Moreover, the Fermi wave length $k_F^{-1}$ of the band near the VHS can satisfy $k_F^{-1} \gg L, \, \xi_K, \, l_s$, where $l_s$ is the length scale over which $V_g(x)$ spatially changes from 0 to $V_g$ at $x=L$. Then, the $2k_F$ term in Eq.~\eqref{scalingbehaviorOfKondotemp} oscillates slowly as a function of $L$ in the range of $L$ where the transition between $T_{K \infty}$ and $T_{K 0}$ occurs; in the case of smooth gate potentials with $k_F l_s \gg 1$, the $2k_F$ oscillations will be washed out.
Finally, to avoid any effects of $V_g$ on $G$ irrelevant to the Kondo effect, one may consider a quantum dot coupled to three wires, applying $V_g$ to one of the wires and measuring $G$ between the other two. 


Note that our system is distinct from the Kondo box (a Kondo impurity in a finite-size system)~\cite{Balseiro,Simon_dot,Hand,Thimm,Bomze}. In the latter, the cloud is terminated hence strongly modified by the box boundary, hence, it is hard to directly detect $\xi_K$. In contrast, in our system, the cloud is extensible to $|x|>L$ (not suppressed even for $L \ll \xi_K$) and can be only weakly (perpurbatively) modified, allowing direct detection of $\xi_K$ and the spatial structure of the cloud.

We thank I. Affleck, G. Finkelstein, L. Glazman, D. Goldharber-Gordon, S. Ilani, A. K. Mitchell, S. Tarucha for useful discussions, and Minchul Lee for advices on NRG calculations. We acknowledge support by NRF (grant 2011-0022955; HSS), and by BSF and Minerva grants (YO). HSS thanks J. Moore and UC Berkeley, where this paper is written, for hospitality.


\begin{thebibliography}{99}

\bibitem{Kondo} J. Kondo, Prog. Theor. Phys. {\bf 32}, 37 (1964).

\bibitem{Hewson} A. C. Hewson, {\it The Kondo Problem to Heavy Fermions} (Cambridge University Press, Cambridge, 1993).

\bibitem{Goldhaber-Gordon} D. Goldhaber-Gordon {\it et al.}, 
Nature {\bf 391}, 156 (1998).

\bibitem{Cronenwett} S. M. Cronenwett, T. H. Oosterkamp, and L. P. Kouwenhoven, Science {\bf 281}, 540 (1998).

\bibitem{kouwenhoven} L. Kouwenhoven and L. I. Glazman, Physics World {\bf 14}, 33 (2001).

\bibitem{Glazman}  L. I. Glazman and M. Pustilnik, in {\it Nanophysics: Coherence and Transport}, eds. H. Bouchiat {\it et al.} (Elsevier, 2005), pp. 427-478. 

\bibitem{Gubernatis} J. E. Gubernatis, J. E. Hirsch, and D. J. Scalapino, Phys. Rev. B {\bf 35}, 8478 (1987). 

\bibitem{Barzykin} V. Barzykin and I. Affleck, Phys. Rev. Lett. {\bf 76}, 4959 (1996); Phys. Rev. B {\bf 57}, 432 (1998).

\bibitem{Borda} L. Borda, Phys. Rev. B {\bf 75}, 041307 (2007). 

\bibitem{Holzner} A. Holzner, I. P. McCulloch, U. Schollw\"{o}ck, J. von Delft, and F. Heidrich-Meisner, Phys. Rev. B {\bf 80}, 205114 (2009). 

\bibitem{Sorensen_knight} E. S. S$\phi$rensen and I. Affleck, Phys. Rev. B {\bf 53}, 9153 (1996). 

\bibitem{Affleck_density} I. Affleck, L. Borda, and H. Saleur, Phys. Rev. B {\bf 77}, 180404 (2008).

\bibitem{Busser} C. A. B\"{u}sser {\it et al.}, Phys. Rev. B {\bf 81}, 045111 (2010). 

\bibitem{Mitchell} A. K. Mitchell, M. Becker, and R. Bulla, Phys. Rev. B {\bf 84} 115120 (2011).

\bibitem{Simon_persistent} I. Affleck and P. Simon, Phys. Rev. Lett. {\bf 86}, 2854 (2001).

\bibitem{Yoshii} R. Yoshii and M. Eto, Phys. Rev. B {\bf 83} 165310 (2011).



\bibitem{Balseiro} P. S. Cornaglia and C. A. Balseiro, Phys. Rev. B {\bf 66}, 115303 (2002); Phys, Rev. Lett. \textbf{90}, 216801 (2003).

\bibitem{Simon_dot} P. Simon and I. Affleck, Phys. Rev. Lett. \textbf{89}, 206602 (2002); Phys. Rev. B \textbf{68}, 115304 (2003). 

\bibitem{Hand} T. Hand, J. Kroha, and H. Monien, Phys. Rev. Lett. {\bf 97}, 136604 (2006). 

\bibitem{Thimm} W. B. Thimm, J. Kroha, and J. von Delft, Phys. Rev. Lett. {\bf 82}, 2143 (1999).

\bibitem {Bomze} Yu. Bomze {\it et. al.}, Phys. Rev. B {\bf 82}, 161411 (2010).

\bibitem{Affleck_review} For a review, see I. Affleck, in 
{\it Perspectives of Mesoscopic Physics} (World Scientific, 2010), pp. 1-44.

\bibitem{Boyce} J. P. Boyce and C. P. Slichter, Phys. Rev. Lett. {\bf 32}, 61 (1974); Phys. Rev. B {\bf 13}, 379 (1976).

\bibitem{Bergmann_G} G. Bergmann, Phys. Rev. B {\bf 77}, 104401 (2008).

\bibitem{Madhavan} V. Madhavan {\it et al.}, Science {\bf 280}, 567 (1998).

\bibitem{Manoharan} H. C. Manoharan, C. P. Lutz, and D. M. Eigler, Nature {\bf 403}, 512 (2000).

\bibitem{Pruser} H. Pr\"{u}ser {\it et al.}, Nat. Phys. {\bf 7}, 203 (2011). 

\bibitem{Fu} Y.-S. Fu {\it et al.}, Phys. Rev. Lett. {\bf 99}, 256601 (2007).

\bibitem{Anderson_Poorman} P. W. Anderson, J. Phys. C {\bf 3}, 2439 (1970).

\bibitem{Wilson} K. G. Wilson, Rev. Mod. Phys. \textbf{47}, 773 (1975); H. R. Krishna-murthy, J. W. Wilkins, and K. G. Wilson, Phys. Rev. B \textbf{21}, 1003 (1980).

\bibitem{Bulla} R. Bulla, T. A. Costi, and T. Pruschke, Rev. Mod. Phys. \textbf{80}, 395 (2008).

\bibitem{Delft} We use the full desnity matrix NRG method developed by A. Weichselbaum and J. von Delft, Phys. Rev. Lett. \textbf{99}, 076402 (2007). We use the NRG discretization parameter of $\Lambda = 2$ and keep $\sim 300$ states at each iteration.

\bibitem{Nozieres} P. Nozi\`{e}res, J. Low Temp. Phys. {\bf 17}, 31 (1974); J. Phys. (Paris) {\bf 39}, 1117 (1978).

\bibitem{Anderson} P. W. Anderson, Phys. Rev. {\bf 124}, 41 (1961).

\bibitem{Schrieffer} J. R. Schrieffer and P. A. Wolff, Phys. Rev. {\bf 149}, 491 (1966). 

\bibitem{Goldhaber-Gordon_Kondotemp} D. Goldhaber-Gordon {\it et al.}, Phys. Rev. Lett. {\bf 81}, 5225 (1998).

\bibitem{Affleck2} I. Affleck and A. W. W. Ludwig, Phys. Rev. B {\bf 48}, 7297 (1993).

\bibitem{Nygard} J. Nygard, D. H. Cobden, and P. E. Lindelof, Nature {\bf 408}, 342 (2000).

\end{thebibliography}
\end{document}